ä \magnification\magstep1
  \baselineskip=14pt
    \font\mybold=cmmib10
   \chardef\Myxi="18
    \def\boldxi{\hbox{\mybold\Myxi}}

   \font\mybold=cmmib10
   \chardef\Myeta="11

   \font\mybold=cmmib10
   \chardef\Myiota="13

   \font\mybold=cmmib10
   \chardef\Mychi="1F
   \def\boldchi{\hbox{\mybold\Mychi}}


      \parindent=0pt

  \centerline{\bf  KILLING TENSORS AND  
CONFORMAL
   KILLING TENSORS}

   \centerline{\bf FROM CONFORMAL KILLING VECTORS}

\

\

\centerline{ Raffaele Rani\dag \footnote{\S}{Present address: Institute of Astronomy and Astrophysics, 
Dept. of Theoretical Astrophysics, University of T\"ubingen,
Auf der Morgenstelle 10 C,
D-72076 T\"ubingen,
Germany.
},   S. Brian
Edgar$\dag$ and Alan Barnes$ \ddag$  }

\centerline{$\dag$ Department of Mathematics, Link\"{o}ping University,
 Link\"{o}ping, Sweden S-581 83.}

\centerline{$\ddag$ School of Engineering \& Applied Science,
Aston University,  Birmingham, B4 7ET, U.K.}

\

E-mail: rani@tat.physik.uni-tuebingen.de, bredg@mai.liu.se, barnesa@aston.ac.uk

\

   {\bf Abstract.}

Koutras has proposed some  methods to construct reducible proper conformal
Killing tensors and  Killing tensors (which are, in general, irreducible)
when a pair of orthogonal conformal
Killing vectors exist in a given space.  We give the completely general result
demonstrating that this severe restriction of orthogonality is
unnecessary. In addition we correct and extend some results concerning
Killing tensors constructed from a single conformal Killing vector. 
A number of examples
 demonstrate how it is possible to construct a much larger class of reducible
proper conformal Killing tensors and  Killing tensors than permitted by the
Koutras algorithms. In particular, by showing that all 
conformal Killing tensors are reducible in conformally flat spaces, we
have a method of constructing all 
conformal Killing tensors and hence all the Killing tensors (which
will in general be irreducible) of conformally flat spaces using  their
conformal Killing vectors.

\

PACS numbers: 0420, 0240

\vfill\eject

{\bf 1. Introduction.}

   A {\it Killing tensor of order 2} is a symmetric tensor $K_{ab}$ such
   that
   $$K_{(ab;c)}=0.\eqno(1) $$
In this paper only Killing and conformal Killing tensors of order $2$ will be considered so in future
this qualification will be assumed tacitly. 
   Physically the interest in Killing tensors is due to their connection
   with quadratic first integrals of geodesic motion and separability
   of classical partial differential equations [1,2,3,4,5,6].

   It is straightforward to show that the metric tensor ${\bf g}$, as
   well as all symmetrised products  of any Killing vectors $\boldxi_I$,
   and, in general, a  linear combination of all of  these with  constant
   coefficients, are all Killing tensors, i.e.,
   $$K_{ab} =  a_0 g_{ab} +\sum^N_{I=1}\sum^N_{J=I} a_{IJ} 
\xi_{I(a}\xi_{|J|b)}\eqno(2)$$
   is a Killing tensor, where $\xi_I$ are the Killing vectors and $a_0$
   and $a_{IJ}$ for $J \ge I$ are constants. Here uppercase
   Latin indices label the  Killing vectors and thus take values
   in the range $1 \ldots N$ where $N$ is the number of independent
   Killing vectors.

   Such Killing tensors  are called  {\it reducible (degenerate,
   redundant} or {\it  trivial)}; all other Killing tensors are called
   {\it irreducible (non-degenerate, non-redundant} or {\it
   non-trivial)}. (Kimura [7] uses 'proper' and 'improper' to
   distinguish between these two classes, but we shall not, since we
   will use those terms in a different context.)

   For $N$ Killing vectors there are in general $N(N+1)/2
   $ symmetrised products of pairs of Killing  vectors, and hence $1 +
N(N+1)/2$ reducible Killing
   tensors; but of course these need not all be linearly independent. In an $n$-dimensional Riemannian
space there exist at most
$n(n+1)/2$ linearly independent Killing vectors, and the maximum
   number can be attained only in  spaces of constant curvature; hence
by substituting $N=n(n+1)/2$ we can obtain the maximum possible number of
reducible Killing tensors in an $n$-dimensional Riemannian space.  
On the other hand, it is known that the maximum number of linearly independent Killing tensors in an
   $n$-dimensional Riemannian space is $n(n+1)^2(n+2)/12$  and the
existence of this maximum number is a necessary and sufficient
   condition for spaces of constant curvature [8,9,10]. So,
for example, in
$4$-dimensional spaces of constant curvature there are a maximum of $10$
linearly independent Killing vectors and hence 
$56 (=1+10.11/2)$
   reducible Killing tensors which can be constructed from the metric and
   the Killing vectors; whereas we know from [8,11] that the
theoretical upper limit of linearly
   independent Killing tensors is only $50(=4.5^2.6/12)$.  
Hauser and Malhiot [12]
   have reconciled these numbers by showing explicitly, in spaces of constant curvature, that of the
   $56$ possible reducible Killing tensors constructed as above, only $50$
are
   linearly independent.   So, in $4$-dimensional spaces of constant
   curvature, all $50$ Killing tensors are reducible.  (This is a
   special case of the more general result [9,13,14,4]  that in
   $n$-dimensional  spaces of constant curvature all Killing tensors are
   reducible.)  

   However, it is the existence of \underbar{irreducible} Killing tensors
   which interests us.  From the physical point of view such tensors
   yield quadratic first integrals which are not simply linear combinations of
   products of the linear first integrals associated with the Killing vectors.  There
   are well known examples of curved spaces which have irreducible
   Killing tensors; for instance, in $4$-dimensional spacetime, the Kerr
   metric [15] has one irreducible Killing tensor [1,2].

  Unfortunately, comparatively few examples of  irreducible Killing tensors
   are known explicitly since the direct integration of $(1)$ is not
   easy, even though there are now computing programs available [16,17,18 and references
therein]. So it would be useful to have
   indirect ways of determining irreducible Killing tensors.

   In this paper we shall consider an indirect method of constructing
   irreducible Killing tensors via  conformal Killing vectors
   which has been proposed by Koutras [19], and also used recently by
   Amery and Maharaj [20].  However, in these two papers the underlying
   principle is not completely transparent nor are the algorithms
   obtained the most general; this is partly due to a distraction caused
   by the trace-free requirement  in the definitions of
conformal Killing tensors which is
   used in these two papers [19,20].  Also in a paper by O'Connor and
   Prince [21] there has been an independent related discussion, but in
   the narrower context of a particular metric.  We shall show that the
   arguments in these papers can be made more general than in the
   original presentations; in particular, we shall show that our more
   general approach enables us to obtain more conformal Killing tensors
   and hence more irreducible Killing tensors than those which can be
   obtained by the algorithms in [19,20].  In addition we shall take the
   opportunity to collect together various results and clarify different
   definitions in the literature.

In Section 2 we establish the basic results, and in Section 3 we
highlight some special cases of these results which are then used for
applications to specific metrics in Section 5.  The results in Section 3
strengthen, extend, and, in one case, correct results in the earlier papers
[19,20]. In Section 4 we  extend a result of Weir [14] for flat spaces
to conformally flat spaces and obtain the maximum number of  conformal
Killing tensors, which shows that they are all reducible in conformally
flat spaces. The results are summarised and further work is discussed in Section 6.

\

\

   {\bf 2. Definitions, Properties and Theorems.}

   We begin with the familiar definitions  in an $n$-dimensional
Riemannian space:

   A {\it conformal Killing vector} $\boldchi$ satisfies
   $\chi_{(a;b)}=\vartheta    g_{ab}$.

   When:

   $\vartheta_{,a}\ne 0, $ $\boldchi$ is  a {\it proper} conformal Killing 
vector.

    $\vartheta=0$, $\boldchi$ is an {\it improper} conformal Killing vector,
which is just a {\it Killing vector}.

$\vartheta_{,a}=0$,
$\boldchi$ is  a {\it homothetic Killing vector}.

   $\vartheta_{,a}=0, \  \vartheta\ne 0$, $\boldchi$ is  a {\it proper}
   homothetic Killing vector.

   \smallskip

   By analogy with the conformal Killing vector, we define:

   A {\it conformal Killing tensor of order 2} is a symmetric tensor
   $Q_{ab}$ such that
   $$Q_{(ab;c)}=q_{(a}g_{bc)}  \eqno(3)$$
   and we easily see that $q_c= {1\over n+2}Q_{,c}+{2\over
n+2}Q^i{}_{c;i}$,
   where $Q=Q^i{}_i$.

   When:

  $q_a=0,$ the conformal Killing tensor $Q_{ab}$ is {\it improper}, and is
 simply a {\it Killing tensor} as defined in (1);

  $q_a\ne 0,$
   the conformal Killing tensor $Q_{ab}$ will be called {\it proper}.

$q_a$ is a Killing vector, $Q_{ab}$ is said to be a {\it homothetic} Killing tensor.  (See Prince [23]
for a discussion of homothetic Killing tensors.)

   The physical interest in proper conformal Killing tensors is due to
   the fact that, although they do not  generate quadratic first
integrals for geodesic motion in general, they do so for null
   geodesics.

   \smallskip

   It is straightforward to show that any scalar multiple of the metric
tensor,
   as well as all symmetrised products of conformal Killing vectors, and,
   in general, all linear combinations of these with constant
   coefficients, are also conformal Killing tensors.

However, a careful consideration of the  definition shows that the number
of linearly independent conformal
   Killing tensors is not finite since, if $ Q_{ab}$ is a conformal
   Killing tensor, then  any other tensor of the form
   $Q_{ab}+\lambda g_{ab}$,
   where $\lambda$ is an arbitrary function of the coordinates
   is also a conformal Killing tensor. To avoid the complication of this
freedom it is usual to subtract off the trace and instead work with
trace-free conformal
   Killing tensors.

A {\it trace-free conformal Killing tensor of order 2} is a symmetric
   trace-free tensor $P_{ab}$ such that $P^i{}_i=0$ and
   $$P_{(ab;c)}=p_{(a}g_{bc)} \eqno(4)$$
   and we easily see that $p_c= {2\over n+2}P^i{}_{c;i}$.

When
$p_a=0,$ the trace-free conformal Killing tensor $P_{ab}$ is {\it improper} (i.e., 
simply a
trace-free Killing tensor), and when
  $p_a\ne 0,$
   the trace-free conformal Killing tensor $P_{ab}$ is {\it
proper}.

   There is no contradiction or ambiguity between the two
definitions. If a trace-free conformal Killing  tensor exists then just by
   adding on an arbitrary trace we obtain a conformal Killing tensor;
   conversely, if a  conformal Killing tensor exists then just by
   subtracting off its trace-term we obtain a trace-free conformal
   Killing tensor.  For a conformal Killing tensor which is 'pure trace'
[24], that is $Q_{ab}=Q g_{ab}/n$, the corresponding trace-free tensor
is identically zero. We also note that, as regards physical interpretation,
the trace part
   does not contribute to the constant of motion along the null geodesics.

   Since we wish to use these properties explicitly, and also to compare
   with earlier results, we will state the most general result as a
   theorem, whose proof can be checked directly.

   {\bf Theorem 1.} In an $n$-dimensional
Riemannian space, if    ${\boldchi_1}, {\boldchi_2}, \ldots , \boldchi_M$ are
conformal
   Killing vectors with associated conformal factors 
$\vartheta_1, \vartheta_2, \ldots , \vartheta_M$, 
any  symmetric tensor of the form
 $$Q_{ab} =  \lambda g_{ab} + \sum^M_{I=1}\sum^M_{J=I}
a_{IJ}
    \chi_{I(a} \chi_{|J|b)}
   \eqno(5a)$$
   is a  conformal Killing tensor with associated vector
$$q_a = \lambda_{,a} + \sum^M_{I=1}\sum^M_{J=I}
a_{IJ}(\vartheta_I \chi_{Ja}+\vartheta_J \chi_{Ia})$$ 
Here  $a_{IJ}$ for $J
\ge I$ are constants; uppercase Latin indices take
   values in the range $1 \ldots M$.
The corresponding trace-free symmetric tensor of the form
 $$P_{ab} =  \sum^M_{I=1}\sum^M_{J=I} a_{IJ}
   ( \chi_{I(a} \chi_{|J|b)}- {1 \over n}\chi_I^{\ c} \chi_{Jc} g_{ab})
   \eqno(5b)$$
   is a trace-free conformal Killing tensor. 
\smallskip
Such  conformal Killing tensors will be called {\it reducible
   (degenerate, redundant} or {\it trivial)}; all other
 conformal Killing
   tensors will be called {\it irreducible (non-degenerate,
   non-redundant} or {\it non-trivial)}.\footnote {${} ^{\dag}$}{Some caution needs 
to be
exercised in reading the literature; in some 
   papers the qualification {\it trace-free} is  not included explicitly in
   the name and so what is sometimes called a 'conformal Killing 
tensor' is in fact 'a trace-free
conformal Killing tensor'.  In this
paper, when appropriate, we shall also retain the
   qualification 'trace-free' to avoid any ambiguities.
 It should also be noted that for {\it reducible conformal
Killing tensors} the factor on the metric is in general non-constant,
unlike for {\it reducible 
Killing tensors} where the factor on the metric must be constant; this
distinction is sometimes blurred in the literature, and in some cases separate definitions
are not even given.}

{\bf Corollary 1.1.}
In a manifold of dimension n>2, if $\boldchi_1$ and $\boldchi_2$ are independent conformal Killing
vectors with associated  conformal factors $\vartheta_1 (\ne 0)$ and $\vartheta_2$ (so that
$\boldchi_1$ at least is  not a Killing vector) then
   $$\chi_{1a} \chi_{1b} \qquad \chi_{1(a} \chi_{|2|b)}$$
   are \underbar{proper} conformal Killing tensors.

{\bf Proof.}
The fact that these tensors are conformal Killing tensors follows immediately from 
Theorem 1.  Hence it remains to show they are proper. Clearly $\chi_{1a} \chi_{1b}$ 
is proper since its associated vector $q_a = \vartheta_1 \chi_{1a}$ is obviously 
non-zero.  The vector $q_a = (\vartheta_1 \chi_{2a} +\vartheta_2 \chi_{1a})/2$ associated with 
 $\chi_{1(a} \chi_{|2|b)}$  cannot vanish since 
otherwise $ \boldchi_2 = -\vartheta_2/\vartheta_1 \boldchi_1$.  However this is 
 not possible since independent conformal Killing vectors cannot be collinear (for dimensions $n>2$).

   \medskip

{\bf Corollary 1.2.}
If $\boldchi_1$ is a proper homothetic Killing vector with associated
conformal factor $h$  and $\boldchi_2$ is a Killing vector then $\chi_{1(a}\chi_{|2|b)}$ is a
homothetic Killing tensor.

The proof is immediate as the associated vector $q_a  = h \chi_{2a}$ is 
clearly a Killing vector since $h$ is a constant and $\boldchi_2$ is a Killing vector.

   \medskip

It is important to note that the total number of reducible conformal Killing tensors as given in 
Theorem
1, will in general be greater than all those reducible conformal tensors obtained by  simply taking
pairs of conformal Killing vectors. Furthermore, it is clear from the corollaries above that a pair of
 conformal Killing vectors, with at least one  proper homothetic or proper conformal, cannot combine
\underbar {directly} to give a Killing tensor; however, the possibility has not been ruled out that
linear combinations of such pairs as in (5a) could give directly $q_a=0$ and hence a Killing tensor.
This possibility is not likely to be common; but we shall now consider a more general possibility of
finding a Killing tensor. 

 \medskip

 A conformal Killing tensor $Q_{ab}$  for which the  vector
$q_a(= {1\over  n+1}Q_{,a}+{2\over n+2}Q^i{}_{a;i})$ is a gradient 
vector, (i.e., $q_a=q_{,a}$),
will be called a {\it conformal Killing tensor of gradient type}. 
It is clear that such a
conformal Killing tensor $Q_{ab}$ will have an {\it
   associated Killing tensor} $K_{ab}$ given by
   $$K_{ab}= Q_{ab}-q g_{ab}\eqno(6a)$$
Such a $K_{ab}$ is defined only up to the addition of a constant multiple of the metric.

If $Q_{ab}$ is a conformal Killing tensor of gradient type then so is 
$Q_{ab} + \lambda g_{ab}$ for any scalar field $\lambda$ and moreover they have the 
same associated Killing tensor.  Thus in particular if $Q_{ab}$ is a conformal Killing 
tensor of gradient type then so is its trace-free part $P_{ab}$ with the same {\it
   associated Killing tensor} $K_{ab}$ given by
   $$K_{ab}= P_{ab}-p g_{ab}\eqno(6b)$$
where  $p_{,a}= {2\over n+2}P^i{}_{a;i}$.

   Walker and Penrose [2] pointed out
   this property for the Kerr metric [15], and  O'Connor and Prince  [21]
have
   exploited this result  for the Kimura metrics [7].   Rosquist and Uggla [22] have  exploited the two
dimensional version of (6b) in investigating a large class of cosmological spacetimes.

   In general, it is the existence of \underbar {irreducible}
 conformal
   Killing tensors that is the most interesting physically. However, in this paper we will only consider 
\underbar
   {reducible proper} conformal Killing tensors, in the expectation that
some of these may be of gradient type and hence will
   yield associated Killing tensors.
   Again, since we wish to use these properties explicitly, and also to
   compare with earlier results, we will state the most general result as
   a theorem, whose proof is direct.

   {\bf Theorem 2.} 
Consider the most general  reducible
conformal Killing
   tensor $Q_{ab}$ of the form (5a).  When there exists a scalar $q$ such
   that
   $$q_{,a} = \lambda_{,a} + \sum^M_{I=1} \sum^M_{J=I} a_{IJ}
   (\vartheta_I \chi_{Ja}+\vartheta_J \chi_{Ia})$$
   then $Q_{ab}$ is a conformal Killing tensor of gradient
type and has
   an associated  Killing tensor $K_{ab}$ where 
   $$K_{ab}=Q_{ab}- q  g_{ab}.$$

\smallskip
    As noted above, the condition of being a gradient conformal Killing tensor is 
unaffected
by
    the addition of an arbitrary trace.  Thus the trace-free qualification is an unnecessary
    complication in the search for Killing tensors associated with reducible conformal
    Killing tensors. We simply construct proper reducible conformal Killing tensors 
   ignoring any considerations of trace, that is setting $\lambda = 0$ in (5a),  and test to see
whether they are of  gradient type and hence yield a Killing tensor.

   If we know all the conformal Killing vectors of a given metric, Theorem 1
   gives us all the reducible conformal Killing tensors as well as the corresponding trace-free reducible conformal Killing tensors, if required. 
Of course if the metric admits $N$ independent Killing vectors $\boldxi_1, \ldots , \boldxi_N$ then the
linear space of all reducible conformal Killing tensors given by (5a) contains a linear subspace of
reducible Killing tensors of the form (2). We can exclude these from consideration if we choose the
basis of the conformal Killing vectors $\boldxi_1, \ldots , \boldxi_N, \boldchi_{N+1}, \ldots
\boldchi_M$ where the $\boldxi_I$'s are Killing vectors and we consider only reducible conformal
Killing tensors of the form 

$$Q_{ab} =  \sum^N_{I=1}\sum^M_{J=N+1} a_{IJ}
    \xi_{I(a} \chi_{|J|b)} + \sum^M_{I=N+1}\sum^M_{J=I} a_{IJ}
    \chi_{I(a} \chi_{|J|b)}
   \eqno(7)$$
where $a_{IJ}$ (for $J\ge I)$ are again constants.

We then test to see if any of these conformal Killing tensors are of gradient type (including
possibility of $q_a=0$) and if so,
 construct the associated Killing tensors.  It is straightforward to check directly
   which of these Killing tensors are irreducible by comparison with equation (2).  
Therefore this is an indirect method to find examples of Killing 
tensors, most of which we expect to be  irreducible, from conformal Killing vectors.

\medskip

   Although the  observations  in Theorems 1 and 2 are very simple, and underlie some 
of the
algorithms given by
   Koutras [19] as well as the generalisations in [20] and  the
   calculation of Killing tensors for the Kimura metric in [21],  
 the argument was not presented so generally or explicitly in these papers. Moreover, in 
[19,20]
there was no explicit definition of  a {\it \underbar{reducible} conformal Killing 
tensor}
(or its trace-free counterpart)  and so the trivial and quite widespread occurence of these tensors
seems to have  been
overlooked. The absence of a means of identifying \underbar{all} reducible conformal 
Killing
tensors meant that \underbar{all} associated Killing tensors could not be found. It 
would
seem that the cause for these less than general results in [19,20] has to do with their
emphasis on \underbar{trace-free} conformal 
      Killing tensors.
They sought  {trace-free} conformal
      Killing tensors constructed as the symmetrised product of a pair of 
     \underbar {orthogonal} conformal Killing vectors, i.e., $A_{i}B^i = 0$, 
      (including the special case of the product of a null Killing vector with itself)
      so that
      $P_{ab} = A_{(a}B_{b)}$
      was automatically trace-free. However, as we have seen a more
      general way to construct a trace-free conformal Killing tensor is
      simply to subtract off the trace from the symmetrised product of two
      conformal Killing vectors, i.e.,  
      $P_{ab}=  A_{(a}B_{b)}-g_{ab}A^{i}B_{i}/n $.

Thus the original results of Koutras [19] on reducible conformal Killing tensors are valid without the
orthogonality  assumptions of his equations (2.3) and (2.9).  We shall show that this more general 
approach enables us to obtain more reducible proper conformal Killing tensors and 
hence more Killing tensors  than those which can obtained by the Koutras  algorithms 
[19].

 \

  \

   {\bf 3 Simple Algorithms for Conformal Killing Tensors and
   Associated Killing Tensors}.

   For a given metric, it will be a straightforward, if long, procedure
   to find the most general conditions on the constants $a_{IJ}$ which
   are required for the existence of a conformal Killing tensor of
   gradient type. However, often there will only be a very limited number
   of possibilities, which can easily be deduced.  So we give some of
   these simpler common  possibilities as corollaries:

   {\bf Corollary 2.1.} The symmetrised product $\xi_{(a}\chi_{b)}$ of a
   Killing vector $\xi_a$ and a proper conformal Killing vector
   $\chi_a$ satisfying $\chi_{(a;b)}= \vartheta g_{ab}$ is a conformal Killing
   tensor of gradient 
type if and only if  $\xi_a$
is a hypersurface
   orthogonal  vector  given by
   $\xi_{a}={ \kappa_{,a}/  \vartheta}$.
   The associated Killing tensor is
   $K_{ab}= \xi_{(a}\chi_{b)} - \kappa g_{ab}$.
   \smallskip
   {\bf Corollary 2.1.1.} The symmetrised product $\xi_{(a}\chi_{b)}$ of a
   Killing vector $\xi_a$ and a proper homothetic Killing vector $\chi_a$
   satisfying $\chi_{(a;b)}= h g_{ab}$, where $h$ is constant, is a homothetic Killing
   tensor of gradient type if and only if
   $\xi_{a}$ is a gradient vector
   $\xi_{a}=  \kappa_{,a}$.
   The associated Killing tensor is
   $K_{ab}= \xi_{(a}\chi_{b)} - \kappa g_{ab}/h$.
   \smallskip
   {\bf Corollary 2.2.} The symmetrised product $\chi_{1(a}\chi_{2 b)}$
   of two different  conformal Killing vectors, respectively satisfying
   $\chi_{1(a;b)}= \vartheta_1 g_{ab}$ and
   $\chi_{2(a;b)}= \vartheta_2 g_{ab}$
   is a conformal Killing tensor of gradient type if and only if
   $\vartheta_2\chi_{1a}+\vartheta_1\chi_{2a}$
   is a gradient vector given by
   $\vartheta_2\chi_{1a}+\vartheta_1\chi_{2a}= \kappa_{,a}$.
   The associated Killing tensor is
   $K_{ab}= \chi_{ 1(a}\chi_{2 b)} - \kappa g_{ab}$.
   \smallskip
   {\bf Corollary 2.2.1.} The symmetrised product $\chi_{1 (a}\chi_{2b)}$ of
   two different proper conformal Killing vectors respectively satisfying
   $\chi_{(1a;b)}= \vartheta_1 g_{ab}$ and
   $\chi_{(2a;b)}= \vartheta_2 g_{ab}$
   which are each hypersurface orthogonal given by
   $\chi_{1a}=  \beta_{,a}/ \vartheta_2$ and
   $\chi_{2a}=  \gamma_{,a}/\vartheta_1$ respectively,
   is a  conformal Killing tensor of gradient type.
   The associated Killing tensor is
   $K_{ab}= \chi_{1(a}\chi_{2 b)} - (\beta + \gamma) g_{ab}$.
   \smallskip

   {\bf Corollary 2.3.} The double product $\chi_{a}\chi_{b}$ of a
   proper conformal Killing vector satisfying
   $\chi_{(a;b)}= \vartheta g_{ab}$
   is a  conformal Killing tensor of gradient type if and only if
   $\chi_a$ is a hypersurface orthogonal vector given by
   $\chi_a= \kappa_{,a}/\vartheta$.
   The associated Killing tensor is
   $K_{ab}= \chi_{a}\chi_{b} - 2 \kappa g_{ab}$.
   \smallskip
   {\bf Corollary 2.3.1.} The  double product $\chi_{a}\chi_{b}$ of a
   proper homothetic Killing vector satisfying $\chi_{(a;b)}= h g_{ab}$,
   where $h$ is constant, is a conformal Killing tensor of gradient type
   if and only if $\chi_{a}$ is a gradient vector field given by
   $\chi_{a}= \kappa_{,a}$.
   The associated Killing tensor is
   $K_{ab}= \chi_{a}\chi_{b} - 2 \kappa g_{ab}/h$.

      \

We can get results for some particular classes of spaces.
   From Corollary 2.1.1 we have directly,

   {\bf Theorem 3.} Any space which admits a proper homothetic Killing
   vector $\chi_a$ with homothetic constant $h$  as well as a gradient
   Killing vector $\xi_{,a}$ also admits a Killing tensor
   $K_{ab}=\chi_{(a} \xi_{,b)}-  \xi g_{ab}/h $.

   \smallskip
  The above  Corollaries 2.1 and 2.3 are given by Koutras [19]  as Theorems 2 and 4 respectively, 
while Theorem 3 above
   is given in [20]; but neither the most general possibility for two  conformal Killing
vectors as given in our Corollary 2.2, nor  the  most
   general result in Theorem 2 above, are given in [19], [20].

\medskip

Exploiting Corollary 2.3 for gradient conformal Killing vectors gives,

   {\bf Theorem 4.} Any space which admits a conformal Killing
   vector field $\chi_a$ which is a gradient also admits the Killing tensor
   $K_{ab}= \chi_a\chi_b - \chi^2 g_{ab}$
   where $\chi^2 = \chi_a \chi^a$.

   {\bf Proof.} As $\chi_a$ is a gradient vector, $\chi_{[a;b]} = 0$,
   and therefore
   $\chi_{a;b} = \vartheta g_{ab}$.
   Thus contracting with $\chi^a$ we
   have $\vartheta \chi_b =  (\chi^2)_{,b}/2$ and the result follows
   from Corollary 2.3.
  \smallskip
  In [19] and [20] it is pointed out that since a geodesic homothetic Killing vector is a gradient,
the result in Corollary 2.3.1 is applicable to such vectors.  Another result for geodesic vectors can be
obtained as follows.

   {\bf Theorem 5.} Any space which admits a proper non-null conformal Killing
   vector field $\chi_a$ which is geodesic (that is
   $\chi_{a;b}\chi^b = \lambda \chi_a$)
   also admits the Killing tensor
   $K_{ab}=\chi_a \chi_b- \chi^2 g_{ab}$.

   {\bf Proof.}
   To see this we contract the equation $\chi_{(a;b)}= \vartheta g_{ab}$
   with $\chi^a \chi^b$ and obtain
   $\lambda \chi^2 = \vartheta \chi^2$. Thus as $\chi^a$ is non-null,
   $\lambda = \vartheta$. Now contracting the equation $\chi_{(a;b)}=
   \vartheta g_{ab}$ with $\chi^b$ and obtain
   $\vartheta \chi_a=(\chi^2)_{,a}/2$ and hence
   $K_{ab}=\chi_a \chi_b -  \chi^2 g_{ab}$ is a Killing tensor.

   \smallskip
   This theorem generalises Theorem 3 of Koutras which was proved in [19] (and also 
quoted in
[20])
   for homothetic Killing vectors only. Our proof is also more direct and
   does not rely on the introduction of a particular coordinate system. In [19] and [20] it
was also claimed to be true in the null case; but it is easy to see that
   the proofs break down in the null case and in fact the result is false as
   the following counter-example shows.  Consider the metric 
$$ds^2 = e^{2u} (2A(x,y,v) du dv +dx^2 + dy^2)$$
A straightforward calculation shows that $\chi^a = \delta^a_u$ is a null homothetic 
Killing vector with conformal factor $\vartheta = 1$.  As $\chi^a$ is a null conformal 
Killing vector it is necessarily geodesic.  The associated conformal Killing tensor 
$Q_{ab}$ and vector $q_a$ are given by
$$Q_{ab} = A^2 e^{4u} \delta^v_a \delta^v_b \qquad q_a = A  e^{2u} \delta^v_a$$
A simple calculation shows $q_a$ is not a gradient vector.

Note also that for non-null vectors Theorem 4 follows from Theorem 5 as a gradient 
conformal
Killing vector is necessarily geodesic.  For the null case a gradient conformal Killing
vector is necessarily a Killing vector and so in this case the associated Killing tensor is
necessarily reducible.

      \

  We emphasise again that in all of these cases the associated Killing tensors
   may or may not be reducible; in each individual space under
   consideration it would be necessary to check directly whether each
   Killing tensor can be reduced to a linear combination of the metric
   and products of pairs of Killing vectors as in equation (2).

\

      {\bf 4. Conformal Transformations.}

      It is well known that, if $\chi^a$ is a conformal Killing vector of the metric
       $g_{ab}$ with confomal factor $\vartheta$ then it is also a conformal Killing
      vector of the conformally related metric $\tilde g_{ab} = e^{2\Omega} g_{ab}$
      with conformal factor
      $\tilde \vartheta = \vartheta + \Omega_{,c}\chi^c$.  We now obtain the analogous
      result for conformal Killing tensors:

      {\bf Theorem 6.} If $Q^{ab}$ is a conformal Killing tensor satisfying
      $\nabla^{(a} Q^{bc)}=q^{(a}g^{bc)}$, then $Q^{ab}$ is also
      a conformal Killing tensor of the conformally related metric 
      $\tilde g_{ab} =  e^{2\Omega}  g_{ab}$.
      $Q^{ab}$ satisfies 
      $\tilde  \nabla^{(a} Q^{bc)}=\tilde q^{(a} \tilde g^{bc)}$,
      where $\tilde q^a = q^a  +2 \Omega_{,d}Q^{da}$.

      {\bf Proof.} The proof is straightforward involving an evaluation of
      $\tilde  \nabla^{(a} Q^{bc)}$ using the result that
       $$\tilde \Gamma^a_{bc} = \Gamma^a_{bc} + \delta^a_b \Omega_{,c}
             + \delta^a_c \Omega_{,b} -\Omega^{,a}g_{bc}$$

 We cannot determine the number of linearly independent conformal
Killing
   tensors because of the freedom in their trace; but we can consider the number of 
linearly 
independent \underbar {trace-free} conformal
Killing tensors.  From Theorem 6 and the analogous result for conformal Killing 
vectors we
have,

{\bf Corollary 6.1.} The number of linearly independent trace-free conformal 
Killing tensors is invariant under conformal change of the metric. The number of 
linearly independent reducible trace-free conformal 
Killing tensors is similarly invariant.

\

   The maximum number of trace-free conformal Killing tensors in an
   $n\ (>2)$-dimensional Riemannian space has
   been found by Weir [14] to be $(n-1)(n+2)(n+3)(n+4)/12$, and he has
   shown that this number is attained in flat space.

   For
   $M$ conformal Killing vectors there are in general
   $M (M+1)/2
   $ symmetrised
      products of pairs of conformal  Killing
      vectors; hence, in conformally flat spaces, we can construct $ M(M+1)/2$
   reducible trace-free conformal Killing tensors.   In an $n$-dimensional Riemannian 
space
 there exist at most $(n+1)(n+2)/2$ linearly
   independent conformal Killing vectors, and the maximum number can be
   attained only in conformally flat
   spaces. Hence by substituting $M=(n+1)(n+2)/2$ we can obtain the maximum 
possible
number of reducible conformal Killing tensors in an $n$-dimensional Riemannian 
space; but of
course these need not all be linearly independent.
     (For example, in
   $4$ dimensions there are
   $120$ reducible trace-free conformal Killing tensors which can be
   constructed from the metric and the
   conformal Killing vectors, while the theoretical upper limit of
   linearly independent trace-free
   conformal Killing tensors is only
   $84$.)
     However, Weir [14]
   has shown explicitly, in $n\ (> 2)$-dimensional flat spaces, that of the $M(M+1)/2$ possible
   trace-free conformal Killing tensors constructed as above, only
   $(n-1)(n+2)(n+3)(n+4)/12$ are linearly independent. So, in
   $n\ (> 2)$-dimensional  flat spaces all
   $(n-1)(n+2)(n+3)(n+4)/12$  trace-free conformal Killing tensors are
   reducible [14].\footnote {${} ^{\dag}$}{Actually Weir only shows that 
   the linear space of trace-free reducible conformal Killing tensors is spanned
   by a certain subset  of $(n-1)(n+2)(n+3)(n+4)/12$ such tensors.
   However it is easy to check that  the
   tensors in this spanning set are linearly independent.} It should be emphasised that these results
do not apply to two dimensional spaces, where Rosquist and Uggla have found some quite different
results   [22].

Applying Corollary 6.1 we can extend Weir's results to conformally flat spaces:

{\bf Corollary 6.2.} The maximum number of linearly independent trace-free conformal 
Killing tensors in $n\ (> 2)$ dimensions is $(n-1)(n+2)(n+3)(n+4)/12$ and is attained in conformally
flat spaces. In this case all the trace-free conformal Killing tensors  are reducible.

   \

      {\bf 5. Examples.}

   The results in the earlier sections are generally valid in $n$
   dimensions.  However,  our main interest
   will be applications in $4$-dimensional spacetime.
   For a given metric with known Killing and conformal Killing vectors, we
first find all \underbar{proper} reducible conformal Killing tensors using Theorem
   1, and can then easily
   find the trace-free versions if required. (We could of course also
write down all reducible Killing tensors, but we concentrate on those
conformal Killing tensors given by (7) that may lead to irreducible Killing tensors.)  
   We could then find all conformal Killing tensors of gradient type, and
      hence all associated Killing tensors via Theorem 2. 
Alternatively we can use the 
      corollaries in Section 3 when the cases are simple.

   \

      \underbar{\it Kimura metric.}

     The Kimura metric (type I in [7], and also considered in [19] and 
[20])  given by
      $$
      ds^{2}={r^{2}\over
      b}dt^{2}-{1\over {r^{2}b^{2}}}dr^{2}-r^{2}d\theta^{2}-r^{2}
      \sin^{2}\theta d\phi^{2}, $$
      is of Petrov type D with a non-zero energy momentum tensor.

      There are  \underbar {four Killing vectors}
      $$\eqalign{\xi_1{}^a= \sin\phi { \partial \over \partial \theta}+ \cot
      \theta
      \cos \phi{ \partial \over \partial\phi} \qquad \hbox{and}\qquad
      \xi_2{}^a= {\partial\over \partial \phi}\qquad \cr
      \xi_3{}^a= - \cos\phi{\partial\over \partial \theta}+ \cot \theta
      \sin \phi{\partial\over\partial \phi} \qquad \hbox{and}\qquad
      \xi_4{}^a= {\partial\over \partial t} \qquad
      }$$
      which in covariant form are
      $$\eqalign{\xi_1{}_a= -r^2\sin\phi \, \theta_{,a}-r^2 \sin
      \theta\cos \theta
      \cos \phi{\, \phi_{,a}} \qquad & \hbox{and}\qquad
      \xi_2{}_a= -r^2 \sin^2 \theta \,\phi_{,a}
      \cr \xi_3{}_a= r^2 \cos\phi{\, \theta_{,a}}-r^2 \sin
      \theta\cos \theta
      \sin \phi \, \phi_{,a} \qquad & \hbox{and}\qquad
      \xi_4{}_a= {r^2\over b}\,{ t_{,a}}}
      $$
       The Killing vectors $\xi_{1a}$ and $\xi_{3a}$ are not hypersurface
      orthogonal, but $\xi_{2a}$ and $\xi_{4a}$ are (but are not gradient).

  There are  also \underbar {two proper
      conformal Killing vectors} with conformal factors $r$ and $rt$
      respectively
      $$
      \chi_{1}{}^{a} = r^2{\partial\over \partial r}
      \qquad \hbox{and}\qquad
      \chi_{2}{}^{a} = r^2 t{\partial\over \partial r}-{1\over
      br}{\partial\over
      \partial t}
      $$
      which are both gradient vectors
      $\chi_{1}{}_{a} = -\Bigl({r / b^2}\Bigl)_{,a}
      \qquad \hbox{and}\qquad
      \chi_{2}{}_{a} = -\Bigl({rt / b^2}\Bigl)_{,a} $ respectively.

   \underbar{Reducible Proper Conformal Killing Tensors}

      From Theorem 1 we can immediately write down $11$ reducible 
proper conformal
      Killing
      tensors from the symmetrised products of each proper conformal Killing
      vector with each Killing vector, together with the symmetrised 
products
      of
      the  proper conformal Killing vectors
 $\xi_{1(a} \chi_{1b)},\ \xi_{1(a} \chi_{2b)},\ \xi_{2(a}
   \chi_{1b)},\ \xi_{2(a} \chi_{2b)},\ \xi_{3(a} \chi_{1b)},\ \xi_{3(a}
   \chi_{2b)},\ \xi_{4(a}
\chi_{1b)},\ \xi_{4(a} \chi_{2b)},$

$\chi_{1a}
   \chi_{1b}, \ \chi_{2a}
   \chi_{2b} ,\ \chi_{1(a}
   \chi_{2b)} \ $;  
it is straightforward
   to find the trace-free versions. These
   $11$ tensors will not
      necessarily
      be linearly independent of each other. 
   \medskip
   \underbar  {Killing Tensors}

        Although the Killing vectors  $\xi_{2a}$ and $\xi_{4a}$ are
   hypersurface orthogonal, neither are
      compatible with the conformal factors $r$ and $rt$ respectively of
   the
      two proper conformal Killing
      vectors to enable Corollary 2.1 to be used.
      On the other hand  $\chi_{1}{}_{a}$ and $\chi_{2}{}_{a}$ are
gradient vectors and so by Theorem 4 we obtain respectively two Killing
tensors with non-zero components, 
   $$\eqalign{& K_{1}^{tt} = {1\over b}\cr
   & K_{1}^{\theta \theta}=  - {1\over b^{2}}     \cr
   & K_{1}^{\phi
   \phi}=  -{1\over b^{2}\sin^{2}{\theta}} }
   $$
  and
   $$\eqalign{& K_{2}^{tt} = b^{2} + {1\over r^{2}} \cr
   & K_{2}^{tr} =-btr                \cr & K_{2}^{\theta \theta}=
   -t^{2}    \cr & K_{2}^{\phi \phi}= - {t^{2}\over
   \sin^{2}{\theta}}.}$$
   Furthermore,  noting that
   $\vartheta_1\chi_{2a}+\vartheta_2\chi_{1a} = - r\Bigl({rt\over
      b^2}\Bigl)_{,a} - rt\Bigl({r\over b^2}\Bigl)_{,a} =
   -\Bigl({r^2t\over
      b^2}\Bigl)_{,a}$ also enables Corollary
      2.2 to be applied to
      $\chi_{1a}$ and $\chi_{2a}$ giving the Killing tensor with
      non-zero components,
   $$\eqalign{& K_{3}^{tt} = 2{t\over b}\cr
   & K_{3}^{tr} = -{r\over b}\cr & K_{3}^{\theta \theta}=-2 {t\over
   b^2} \cr & K_{3}^{\phi \phi}=  - {t^{2}\over \sin^{2}{\theta}}
   b^{2} .}
   $$
   To check whether there may be more, less obvious Killing tensors,
   we apply Theorem 2 in the most general case and consider the
   vector
   $$\eqalign{
   a_{55} r &(r^2/b^2)_{,a}+a_{66} rt (r t/b^2)_{,a}+(a_{15}r+a_{16}r
   t)(-r^2\sin\phi \, \theta_{,a}-r^2 \sin
      \theta\cos \theta
      \cos \phi{\, \phi_{,a}})\cr &+(a_{25}r+a_{26}r
   t)(-r^2 \sin^2 \theta \,\phi_{,a}   )  +(a_{35}r+a_{36}r t)(r^2
   \cos\phi{\, \theta_{,a}}-r^2 \sin
      \theta\cos \theta
      \sin \phi \, \phi_{,a} )\cr & \qquad 
   +(a_{45}r+a_{46}r t)({r^2\over b}\,{ t_{,a}} ) +a_{56}(rt
   (r^2/b^2)_{,a}+ r (r t/b^2)_{,a})}$$ It is immediately obvious
   that for this vector to be gradient, all of
   $a_{15},a_{25},a_{35},a_{16},a_{26},a_{36}$ must be zero. From the
   remainder we find the gradient condition is equivalent to
   $$
   0= (a_{45}\, r_{,[b}+a_{46} t \,\, r_{,[b})({ t_{,a]}} )
   $$
   and so
   $a_{45}$ and  $a_{46}$ must also be zero.  Therefore, the 3
   Killing tensors found above are the only ones which can be
   obtained by this method.

   A comparison of the Killing tensor ${\bf K}_1$ with the Killing
   vectors shows that it is in fact reducible since, [19]
   $$K_{1ab}= {1\over b} \xi_{4a} \xi_{4b} - {1\over b^2} ( \xi_{1a}
   \xi_{2b}+\xi_{2a} \xi_{2b}+\xi_{3a} \xi_{3b})  .$$
   The
      other two Killing tensors are  irreducible since it is clearly
   impossible to
   obtain, using the Killing vectors and metric,  those
      terms in $K_{2ab}$ and $K_{3ab}$ which are explicit functions of
   $t$. It is easy to confirm from observation that these three 
tensors are linearly independent of each other
and of the metric.
   \smallskip

   (In Kimura's [7] original work he sought directly for irreducible
   Killing tensors, and found  $\bf{K_2} $ and $\bf{K_3}$. Koutras
   [19] only found 8 reducible trace-free conformal Killing tensors
   and
       only the 2 Killing tensors $\bf{K_1}$ and $\bf{K_2}$,  because
   he
   used his less general
      algorithms. However, O'Connor and Prince [21] obtained all three
   Killing tensors since they used the same more general  argument as
   we have done.)

   \medskip

   \underbar{\it  Bell-Szekeres metric}

   We now consider the Bell-Szekeres metric [25], a Petrov type I metric, which in coordinates
   $(u,v,x,y)$ is defined by the line element
   $$
   ds^2 = 2(u+v)^{(a^2-1)/2}dxdv -2(u+v)^{1-a}dx^2 -2(u+v)^{1+a}dy^2
   $$
   where $a$ is a constant; when $a=0,\pm 3$ the space is Petrov type 
D and when $a= \pm 1$ the
space is flat. We concentrate on curved spaces.

   It admits the \underbar{three Killing vectors}
   $$\eqalign{
   & \xi_{1}^{a}= {\partial\over \partial u}-{\partial\over \partial
   v},\cr & \xi_{2}^{a} ={\partial\over \partial x}, \cr &
   \xi_{3}^{a}= {\partial\over \partial y} , }$$
   which are all hypersurface orthogonal
   $$
   \eqalign{& \xi_{1a}= (u+v)^{{(a^2-1)\over 2}} (v-u)_{,a} \cr &
   \xi_{2a} =-2(u+v)^{(1-a)}x_{,a} \cr & \xi_{3a}=
   -2(u+v)^{(a+1)}y_{,a},}$$
None of these three vectors, nor any combination, can be gradients 
(except in flat space).

  Furthermore, this metric also possesses \underbar{one proper homothetic
   Killing vector}, [26]
   $$ \chi^a = 4u {\partial\over \partial u} + 4v
   {\partial\over \partial v} + (1+a)^2x {\partial\over \partial x}
   +(1-a)^2y {\partial\over \partial y} $$
   with conformal factor $3+
   a^2$; it is easy to see that this vector is not gradient. 

\smallskip
   \underbar{Reducible Proper Conformal Killing
   tensors}

      From Theorem 1 we can immediately write down $3$ reducible proper
   homothetic
      Killing
      tensors from the symmetrised products of the proper homothetic
   Killing
      vector with each Killing vector; in addition we have one reducible proper
   conformal
      Killing from the double product
      of
      the homothetic Killing vector:
$\xi_{1(a} \chi_{b)}, \  \xi_{2(a} \chi_{b)},\ \xi_{3(a}
   \chi_{b)},\ \chi_{a} \chi_{b} $;  it is straightforward to find
   the trace-free versions by subtracting off the traces. Since the proper homothetic vector is not
null, nor  orthogonal to any of the Killing vectors, these four  tensors could not  have
been obtained from the Koutras algorithms. 
   \smallskip

   \underbar  {Killing Tensors}

  The fact that neither the homothetic Killing vector  nor the 
Killing vectors are gradient
vectors means that via Corollaries 2.1.1 and 2.3.1 we can conclude 
that  we   cannot
   construct
   any Killing tensors by this method.

  \medskip
   \underbar{\it Beem metric }

   The Beem metric [27] is defined in coordinates $(u,v,x,y)$ by the
   line element
   $$
   ds^2= {e^{vx}}dvdu+dx^2+dy^2 $$ and possesses \underbar{two Killing
   vectors}
   $$
   \eqalign{ &\xi_1^a = {\partial\over \partial u} \cr &\xi_2^1=
   {\partial\over \partial y} }
   $$
   of which $\xi_{1a}$ is hypersurface orthogonal $$\xi_{1a}=
   {1\over 2}{e^{vx}} v_{,a} $$
   and $\xi_{2a}$ is gradient
   $$
   \xi_{2a} = y_{,a}. $$ Furthermore, this metric admits \underbar {one proper
   homothetic vector}
   $$
   \chi^a = 3u {\partial\over \partial u} -v {\partial\over \partial
   v}+ x{\partial\over \partial x} +y{\partial\over \partial y}
   $$
   with homothetic factor 1; this vector is not a gradient.
   \smallskip
   \underbar {Reducible Proper Conformal Killing tensors}

   We can construct two proper homothetic Killing tensors and one proper conformal Killing tensor by
using
   $\chi_a$ with $\xi_{1a}$ and $\xi_{2a}$
   respectively, as well as $\chi_{a}\chi_{b}$. Again these could not have been obtained using the
Koutras algorithms.
   \smallskip
   \underbar {Killing tensors }

The homothetic vector together
   with a gradient Killing vector can be used to construct a Killing
   tensor according to Corollary {2:1:1}, and we obtain
   $$\eqalign{
   &K^{uv} =-2 {y {e^{-vx}}}\cr &K^{uy} ={3\over 2}u \cr &K^{vy}
   = -{1\over 2}v  \cr &K^{xy} = {1\over 2}x   \cr &K^{xx} = -y }
   $$
   Checking with the Killing vectors and metric we note that this
   Killing tensor is irreducible.
We cannot obtain any more Killing tensors from Theorem
   2.

\medskip

      \underbar{\it  Fluids   with gradient conformal Killing vectors} 

   In [30]  examples of perfect and other  fluid spacetimes  which
   admit a gradient conformal Killing
      vector are found; for all such spaces Killing tensors can be constructed using Theorem 4. 
In [20] the possibility of such a construction was pointed out --- but  only for any of these 
spaces
whose gradient conformal Killing vector is homothetic. 

\medskip

\underbar{\it  Conformally flat spaces}

Conformally flat spacetimes necessarily admit 15 independent conformal Killing vectors from which 
84 independent reducible conformal Killing tensors can be constructed.  Hence in such spacetimes there
is a rich supply of 'candidate' conformal Killing tensors which may satisfy the gradient condition and
so be associated with possibly irreducible Killing tensors.  The large number of candidate tensors
means that a direct approach by hand calculation would be lengthy and error-prone.  However the
calculations involved, though lengthy, are routine and this enables them to be automated by using of a
computer algebra package such as Reduce.  Work is in progress on investigating a number of conformally
flat spacetimes including the  perfect fluid solutions [28] and the   pure
radiation solutions [29], the Robertson-Walker metrics and the interior Schwarzschild solution; 
these results will be presented elsewhere.  In this paper we will restrict ourselves to a few
preliminary remarks.

The generic perfect fluid solutions [28] and the   pure
radiation solutions [29]   admit no Killing vectors and so if any gradient conformal Killing tensors
are found then the associated Killing tensors will necessarily be irreducible (unless they are simply
constant multiples of the metric).

Amery and Maharaj [20] found a number of conformal Killing tensors and Killing tensors in
Robertson-Walker spacetimes using Koutras' algorithms, but because they used only mutually orthogonal
conformal Killing vectors in their construction, they were only able to construct 39 'candidate'
conformal Killing tensors.  However, the Robertson-Walker metrics, being conformally flat, admit the
maximal number, namely 84, of reducible conformal Killing tensors and so Amery and Maharaj's results
are incomplete.

A generic Robertson-Walker metric admits 6 independent Killing vectors and so 
22 (= 1 + 6.7/2) reducible Killing tensors can be constructed from the metric and the Killing vectors
--- of which 21 are linearly independent.  Similarly for the special case of the static Einstein
universe which admits a seventh Killing vector, we can construct 30 (=1+7.8/2) reducible Killing
tensors  from the metric and the Killing vectors --- of which 27 are linearly
independent.    Hence, after finding the gradient conformal Killing tensors and their associated Killing
tensors of the generic Robertson-Walker metric (or of the Einstein universe), we need to determine
whether they are irreducible by checking if they are independent of these 21 (or 27) reducible Killing
tensors.  Again the high dimension of these linear subspaces involved and the routine nature of the
calculations means that the computations can be automated by use of the computer algebra system Reduce.

   \

   {\bf 6. Discussion}

  We have clarified the concept and definition  of {\it reducible}  conformal Killing
tensors of order 2 and their trace-free counterparts; this enables us to write down immediately all
the reducible conformal Killing tensors in a space where the conformal Killing vectors 
are
known. By identifying those reducible  conformal Killing
tensors of gradient type we are able to construct associated Killing tensors, most of 
which
we expect to be irreducible. For conformally flat spaces we have shown that all 
conformal
Killing vectors are reducible and so they can all (including both reducible and 
irreducible
Killing tensors) be found by this indirect method.

 Of course,   in more general curved spaces there are important examples of
   irreducible trace-free conformal Killing
   tensors, and also of irreducible Killing tensors which are not associated 
 with conformal
   Killing vectors; such tensors cannot be obtained by the indirect method in this paper. 
However there are other possibilities of indirect methods which may enable us to find 
some
of these other Killing tensors. For example,  we can construct Killing tensors  from
Killing-Yano tensors, and conformal Killing tensors  from conformal Killing-Yano 
tensors;
we could check which of these are reducible in the sense of the respective definitions 
given
in this paper. Moreover,
any  proper conformal Killing tensors which are
of gradient type would then give associated Killing tensors, which in general would not 
be
constructed  from Killing vectors, or even from Killing-Yano tensors.

      \

      \

      {\bf Acknowledgements.}  B.E. wishes to thank Hans Lundmark for discussions
and references, and Vetenskapsr\aa det (the Swedish Research Council) for financial support. Both
authors thank Kjell Rosquist for drawing attention to the reference to the two dimensional case, and
its difference from the general $n\ (> 2)$-dimensional case.

\medskip
{\bf References}
\smallskip

1.     B. Carter, {\it Phys. Rev.}, {\bf 174}, 1559 \ (1968).
\smallskip

2.   M. Walker and R. Penrose, {\it Commun. Math. Phys.}, {\bf 18}, 265  (1970).

\smallskip

3. N.H.J. Woodhouse, {\it Commun. Math. Phys.}, {\bf 44}, 9  (1975).

\smallskip
4.  S. Benenti, {\it J. Math. Phys.}, {\bf 38}, 6578 (1997).
\smallskip
5.  S. Benenti and M. Francaviglia,  {\it General Relativity and Gravitation Vol.1,} p. 393.   Ed. A.
Held, Plenum Press, New York (1979).
\smallskip

6.  E. Kalnins and W. Miller, {\it SIAM J. Anal.}, {\bf 11}, 1011 \ (1980).

\smallskip
7.     M. Kimura,   {\it Tensor N.S.}, {\bf 30}, 27 \ (1976).
\smallskip
8.      T.Y. Thomas, {\it Proc. N. A. S.}, {\bf 32}, 10 \ (1946).

\smallskip
9.     G. H. Katzin and I. Levine, {\it Tensor N.S.}, {\bf 16}, 97 \
   (1965).

\smallskip

10.    C.D. Collinson, J. Phys. A: Gen. Phys., {\bf 4}, 756 \ (1971).
\smallskip
11.  P. Sommers  {\it J. Math. Phys.}, {\bf 14}, 787 (1973).

\smallskip
12.     I. Hauser  and R.J.  Malhiot, {\it J. Math. Phys.}, {\bf 16},
      1625 \  (1975).

\smallskip
13.  G. Thompson,  {\it J. Math. Phys.}, {\bf 27}, 2693 (1986).
\smallskip
14.   G. J. Weir, {\it J. Math. Phys.}, {\bf 18}, 1782 (1977).
\smallskip

15.     R. P. Kerr,   {\it  Phys. Rev. Lett.}, {\bf 11},  237 (1963).
\smallskip

16. G.C.  Joly,  {\it Gen. Rel. Grav.}, {\bf 19}, 841 (1987).

\smallskip
17.      A. Koutras  and J.E.F.  Skea,  {\it Computer Physics
      Communications}, {\bf 115}, 350 \ (1998).
\smallskip
18.   T. Wolf,  {\it Gen. Rel. Grav.}, {\bf 30} 124 (1998).

\smallskip
19.     A. Koutras, {\it Class. Quantum Grav.}, {\bf 9}, 1573 \ (1992).
\smallskip

20.      G. Amery and S.D Maharaj, {\it Int. J. Mod. Phys. D}, {\bf 11},
      337-351 \ (2002).
\smallskip
21.        J.E.R.  O'Connor and G.E. Prince,
      {\it Class. Quantum Grav}. {\bf 16}, 2885 \ (1999).  
\smallskip
22. K. Rosquist and C. Uggla,  {\it J. Math. Phys.}, {\bf 32}, 3412 (1991).

\smallskip
23.   G. Prince, {\it Physics Letters}, {\bf 97A}, 133 \ (1983).

\smallskip

24. R. Geroch {\it Commun. Math. Phys.}, {\bf 13}, 180 \ (1969).

\smallskip
25.  P. Bell and P. Szekeres, {\it Int. J. Theor.
   Phys.}, {\bf 6},  {111}, (1972).

\smallskip

26. S.B. Edgar  and G. Ludwig,  {\it
Gen. Rel. Grav.}, {\bf 34}, 807 \ (2002).

 \smallskip
27.   J. K. Beem,  {\it Letters Math. Phys.},  {\bf 2}, 317 \ (1978).

\smallskip
28.   H. Stephani, {\it Commun Math. Phys.}, {\bf 4}, 137 \ (1967).
\smallskip
29.   G. Ludwig and   S. B. Edgar, {\it Class. Quantum Grav.}, {\bf 14}, L47 \ (1997).
\smallskip
30.  V. Daftardar and N. Dadhich, {\it
Gen. Rel. Grav.}, {\bf 26}, 859 \ (1994).

\end